%% file: make_astro.tex
\documentclass{mpe_report}

\usepackage{psfig,graphicx,epsfig}
\usepackage{color}
\usepackage{amsmath,amssymb,epic,eepic,array}

\begin{document}

\pagenumbering{arabic}
\setcounter{page}{223}

 \renewcommand{\FirstPageOfPaper }{223}\renewcommand{\LastPageOfPaper }{225}\include{./mpe_report_keith}              \clearpage

\end{document}

%% file: mpe_report_keith.tex
\title{Pulsar Virtual Observatory}
\author{M. Keith\inst{1} \and B. Harbulot\inst{2} \and A. Lyne\inst{1} \and J. Brooke\inst{2}}  
\institute{Jodrell Bank Observatory, University Of Manchester, Macclesfield, Cheshire SK11 9DL, United Kingdom
\and  School of Computer Science, University of Manchester, Oxford Road, Manchester M13 9PL, United Kingdom}
\maketitle

\begin{abstract}
The Pulsar Virtual Observatory will provide a means for scientists in all fields to access
and analyse the large data sets stored in pulsar surveys without specific knowledge about the
data or the processing mechanisms.
This is achieved by moving the data and processing tools to a grid resource where the
details of the processing are seen by the users as abstract tasks.
By developing intelligent scheduling middle-ware the issues of interconnecting tasks and
allocating resources are removed from the user domain.
This opens up large sets of radio time-series data to a wider audience, enabling greater
cross field astronomy, in line with the virtual observatory concept.
Implementation of the Pulsar Virtual Observatory is underway,
utilising the UK National Grid Service as the principal grid resource.

\end{abstract}

\section{Background}

The concept of a Virtual Observatory (VO) is to enable access to astronomical data archives via a generic interface \cite{quin:2004}.
This interface can be some well specified machine interface, e.g. an XML
schema, or a direct to user interface, e.g. a HTML interface.
The use of a generic interface is to reduce the level of domain specific knowledge about the data, enabling users from a wide range of backgrounds to access the data.
This reduces the issues involved in extracting and processing data from many resources, allowing scientists to search for answers across a wide range of data archives.
Standardisation of communication protocols allows for intercommunication between many VO projects, further widening the scope of research.

The Pulsar Virtual Observatory aims to apply the VO concept to data from archived pulsar surveys.
This will enable new pulsar science as astronomers and theoreticians can access data that would otherwise be hidden.
Access will be made as simple as possible by providing a web interface that can be used by a standard web browser application.
There will also be a range of options to allow users with different levels of experience to benefit fully from the Pulsar Virtual Observatory.

\section{Data archives}
The Pulsar Virtual Observatory has not been designed with any constraints over the data format that is used, and it is intended that processing can cover multiple data formats seamlessly.
Data will be indexed by sky position as it is expected that users will search data at specific locations of interest, and so access to multiple data catalogues that cover the same sky area is advantageous.

Initially the data set that will be made available via the Pulsar Virtual Observatory will be the Parkes Multi-beam Pulsar Survey (PMPS) raw data archive \cite{manchester:2001}.
This is the most complete survey of the galactic plane to date, and provides an ideal resource for finding new pulsars and other transient radio phenomenon.
Data was collected using the $13$ beam HI (21cm) receiver on the 64m Parkes telescope in Australia and quickly became the most successful pulsar survey to date.
The archived data covers the galactic plane from $260^0$ to $50^0$ in galactic longitude and $\pm5^0$ of galactic latitude.
The raw data archive is around $4.4$ terabytes in size, consisting of approximately $35000$ integrations of $35$ minutes with a $288$ MHz band centred on $1374$ MHz.
The data archive has been extensively searched for period signals, however more pulsars continue to be discovered with in depth targeted searches and by using more advanced search tools\cite{faulkner:2004}.

This data is stored as raw data, as it is taken from the telescope, giving the maximum flexibility in analysing the data.
Working with raw data means that the analysis can be performed with the latest algorithms, which are constantly being improved.
This allows for expansion of the functionality of the Pulsar Virtual Observatory that would not be possible with archives of search results, or partially processed files.
The Pulsar Virtual Observatory can also be expanded by adding new data archives to the system.
This will enable new types of search, e.g. searching in different frequency bands, as well as expanding the sky coverage available.
It is expected that more Parkes surveys can be added to the system when the data is made publicly available.

There are some issues regarding data storage that need to be addressed when developing this system.
Currently the raw data archives are stored on RAID disks at Jodrell Bank Observatory.
It is intended that these archives be made available to the Pulsar Virtual Observatory via a simple file transfer interface, however most frequently accessed data can be moved close to the processing nodes, to reduce data
transfer times.
The design of the Pulsar Virtual Observatory makes it possible to store data in multiple locations, allowing for replication of data for backup or speed purposes.
It is intended that the system also implement a pre-fetching system, where the data is transfered to a location near to the processing node prior to the processing starting.
This is can be achieved by transferring the appropriate data file when the task is entered into a processing queue.
This reduces the processor time wasted while waiting for the file to transfer to the node that is performing the computation.

\section{Computation resources}
The Pulsar Virtual Observatory will provide access to compute resources as well as the large data archives.
Processing will be made available via a simple web interface, directly linked to the searchable data archive.
This design allows users to analyse data without the need to locate and install software to their local systems.
This enables greater access to the data as many users do not have the knowledge, skills, resources or time required to install and run the appropriate software.
Advanced users can still be given the opportunity to customise the software that is being used, or download the data to process by their own means.

The current target system is the UK National Grid Service\footnote{http://www.ngs.ac.uk} (NGS) a grid computing resource for UK science projects.
There is however nothing to prevent the Pulsar Virtual Observatory using other resources in combination with or instead of the NGS.

The NGS is a dedicated high performance grid resource for the UK eScience programme.
It comprises four core nodes at Manchester, Oxford and Leeds universities and at the Rutherford Appleton Laboratories.
The NGS is suitable to be used as our initial target system as it provides high performance computing and data storage facilities in a grid environment.
The use of grid facilities makes integration with the Pulsar Virtual Observatory easy, as each site can be accessed via a uniform interface.

Because the Pulsar Virtual Observatory can use multiple resources for analysing data, the system is designed in such a way that the user does not need to know about the system their jobs are running on.
This means that users of the Pulsar Virtual Observatory will not require separate authentication for each system that processing tasks are to be run on.
By providing a simple user interface, it is possible for users from outside the immediate scientific field to analyse data, leading to more cross-field science, in line with the VO principles.
It is intended that there will be direct access to the raw data if required, allowing advanced users to develop their own processing algorithms.

Telescope data can be searched with user specified parameters, using a supplied set of search tools.
For example, a user could select a number of telescope pointings and perform a Fourier analysis on a time-series generated by the dedispersion of these pointings.
Users will be able to perform other types of searches, and new algorithms will be added when they are made available.

\begin{figure}[ht]
\centerline{\psfig{file=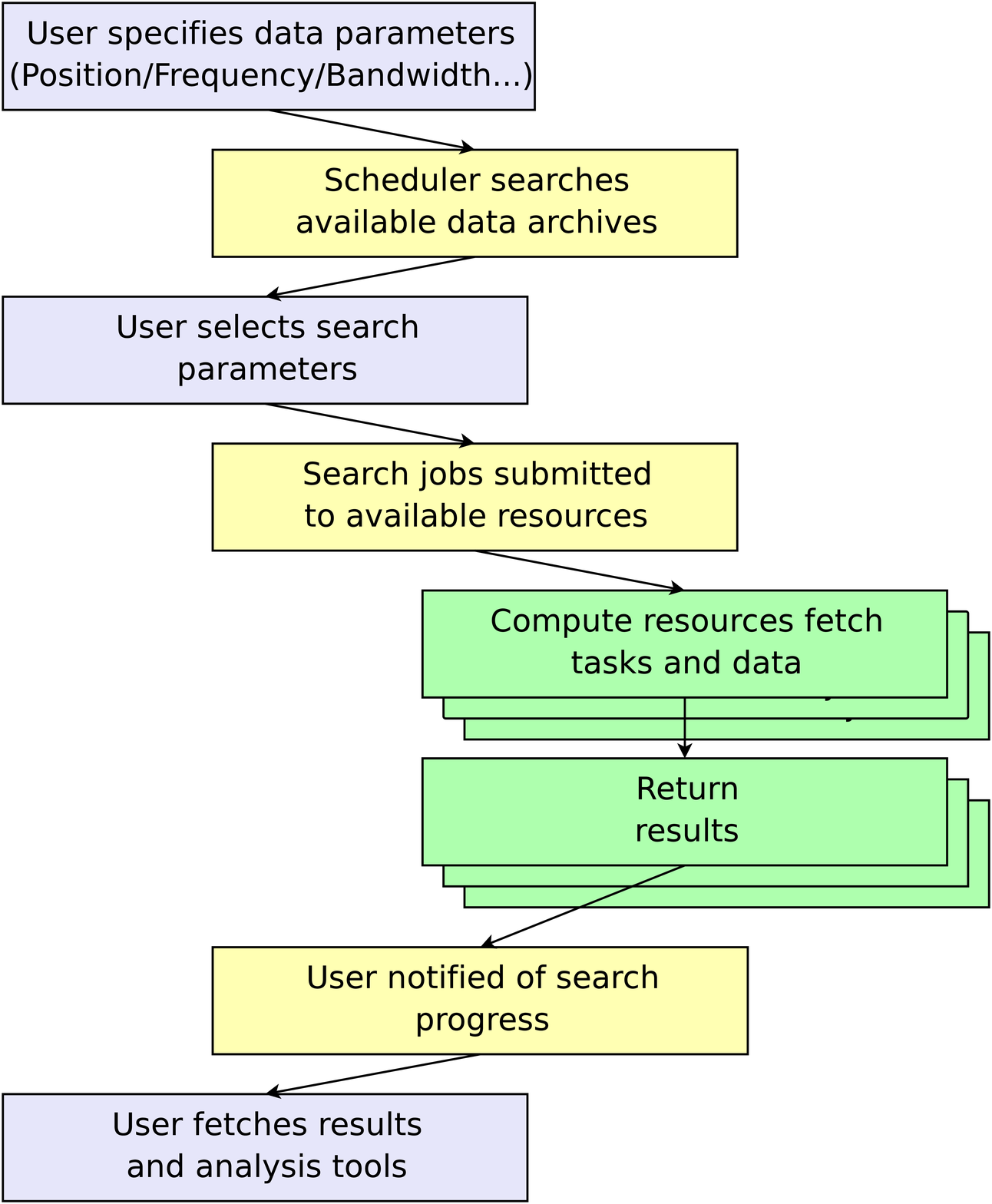,width=8.8cm,clip=} }
\caption[A simple overview of the search process on the Pulsar Virtual Observatory.]{
A simple overview of the search process on the Pulsar Virtual Observatory. 
\label{workflow}}
\end{figure}

The search parameters are user customisable, however pre-defined default values will also be available for many options.
This means that expert users will be able to customise the search routines for their specific task, but still giving wide access to users with less experience.
The system will also provide feedback regarding the status and remaining runtime of tasks.
This means that users will be able to better optimise the settings for their jobs.

\section{Implementation}

The core architecture of the Pulsar Virtual Observatory is based around a three layer design.

The top layer handles user interaction and generates the tasks that are required to process the user-selected data.
It is also responsible for scheduling the tasks on the available compute resources.

Compute resources are modelled by a web services based queueing system, developed in parallel with the Pulsar Virtual Observatory \cite{harbulot:2006}.
The queueing system is a application independent, i.e. there is no code that is specific to the Pulsar Virtual Observatory, and is platform independent, i.e. there is no code specific to the compute
resources that the tasks are being run on.
By having tasks pulled from the queues, rather than pushed onto the processing resources, the queueing system does not need any specialist code to interface with particular resources.
This means that all resources, whether single machines, clusters or grids are viewed through the same interface.

The low level computation is handled by the final layer, which is handled by a series of simple shell scripts.
This approach allows for deployment on a wide range of systems as it does not rely on any specialist software being installed.
As tasks are pulled from the queue system there is no need to run the scripts in a listening mode.
This means there is no need for inbound communication, which is often restricted, and the queueing server does not need to know the physical location of the machine that is running the processing script.

User interface is provided via a web interface that communicates directly with the scheduler a database of the available data archives and current process status.
This will provide abilities to search the data catalogues and submit new jobs, as well as monitor progress of running jobs.
The web interface will also provide tools for viewing the results of the data analysis, although the user may download the results to analyse with their own software.

\section{Current status}
The Pulsar Virtual Observatory is not yet fully operational, however a large fraction of the implementation is complete.

There is a searchable database of the data available in the system, which is accessible via the web interface.
Users can select this data and select and customise processing to be performed on the data.
This work is then scheduled on the available processing nodes via a simple scheduling algorithm.
Processing algorithms have also been successfully deployed on the NGS, and test workflows have been completed.

Many issues remain to be resolved however, including error handling for failed jobs, user notification for job status updates and remaining runtime estimation.
There is still a lot of development work that is required to get the output from the processing software back to the user, and to provide satisfactory visualisation tools.
Data pre-fetching is not yet implemented.

Testing of the basic system is currently in progress, and it is expected that the system will soon be usable for a selection of trial problems.

\section{Conclusion}
The Pulsar Virtual Observatory will provide access to data and resources in order to enable greater use of archived pulsar survey data.
This will allow users to analyse pulsar data in a simple and uniform manner, without detailed knowledge of data formats and processing software, encouraging more science to be carried out.
The Pulsar Virtual Observatory is designed to integrate with other VO projects and so enable integration of pulsar data archives with data sets from other astronomical fields.

%\begin{figure}[h]
%\centerline{\psfig{file=$hs/Keith/workflowsimple2.eps,width=8.8cm,clip=} }
%\caption{Example of an included figure in a column( .ps file).
%\label{image}}
%\end{figure}

\begin{acknowledgements}
Michael Keith is funded by the Particle Physics and Astronomy Research Council (PPARC).

Bruno Harbulot acknowledges funding under PPARC Project PP/000653/1 (GridOneD).
\end{acknowledgements}
   
%\bibliography{mn-jour,mkeith} 

%\end{document}